\documentclass[prb,twocolumn,floatfix,superscriptaddress]{revtex4} 
\usepackage{graphicx,natbib,epsfig,bm}
 \newcommand{\bea}{\begin{eqnarray}}
 \newcommand{\eea}{\end{eqnarray}}
 \newcommand{\pdag}{{\phantom{\dagger}}}
\begin{document}
\title{Singlet--Triplet Transition in lateral Quantum Dots: \\
A Numerical Renormalization Group Study}
\author{W. Hofstetter}
\email{hofstett@cmt.harvard.edu}
\affiliation{Lyman Laboratory, Harvard University, Cambridge, MA 02138, USA}
\author{G. Zarand}
\email{zarand@newton.phy.bme.hu}
\affiliation{Theoretical Physics Department, Budapest University of Technology and Economics, Budafoki \'ut 8. H-1521 Hungary}
\affiliation{Lyman Laboratory, Harvard University, Cambridge, MA 02138, USA}

%\date{\today}

\begin{abstract}
We discuss transport through a lateral quantum dot in the vicinity of a singlet-triplet spin 
transition in its ground state. Extracting the scattering phase shifts from the  numerical renormalization group
spectra, we  determine the linear conductance at zero temperature as a function of a Zeeman 
field and the splitting of the singlet and triplet states.  We find reduced low-energy transport,
and a non-monotonic magnetic field dependence  both in the singlet 
and the triplet regime. For a generic set of dot parameters and no Zeeman splitting, 
the singlet-triplet  transition may be identified with the conductance 
maximum. The conductance is least sensitive to the magnetic field 
in the region of the transition, where it decreases upon application 
of a magnetic field. Our results are in good agreement with recent 
experimental data.
\end{abstract}

\maketitle

\section{Introduction}

Recent observations of the Kondo effect in semiconductor quantum dots 
\cite{kondo,kondo-theo}  induced  enormous theoretical and experimental activity.
 Compared to magnetic impurities in solids\cite{Hewson},  quantum 
 dots\cite{Coulomb-blockade}  
have the great advantage of tunable couplings, and as a result, 
in these systems the full parameter regime of the single-level Anderson 
impurity 
model can  be systematically explored. 
In addition, new types of Kondo systems have been realized, 
like Aharonov--Bohm rings containing quantum dots\cite{Ji00}  
and multi--level dots\cite{Tarucha00}. 
In vertical dots, a qualitatively new type of Kondo effect
associated with a  singlet--triplet degeneracy  
has been observed \cite{Sasaki00},  
which was later also explained theoretically\cite{Eto00,PG1,PG2}.

For lateral dots, this \emph{singlet--triplet} Kondo effect 
has been found as well\cite{vanderWiel02}, but with qualitatively new behavior 
at low energies, where both the linear conductance and $dI/dV$ 
were found to be \emph{non--monotonic}. 
Theoretical understanding of this behavior has been obtained 
in two limiting cases: Hofstetter and Schoeller\cite{Hofstetter02} 
have analyzed the interplay between singlet and triplet configurations 
with \emph{symmetric} coupling to the leads, while 
Pustilnik and Glazman\cite{real} have considered 
general asymmetric couplings but took only the triplet configuration 
on the dot into account. 

Agreement with the singlet--triplet scenario of 
[\onlinecite{Hofstetter02}] 
has recently been found in another experiment\cite{Kogan02}, 
where -- instead of the magnetic field as in [\onlinecite{vanderWiel02}] --
the Stark effect due to the gate potential was used to 
tune the level splitting. Another way to control the level spacing 
and to study the triplet-singlet transition  by  a symmetric (triangular) 
arrangement has been proposed in [\onlinecite{David}].

Our goal in this work is to extend the studies of 
[\onlinecite{Hofstetter02,real}] over the entire range of parameters
in a non-perturbative way, and to include singlet/triplet 
degeneracy as well as asymmetry of the tunnel couplings in our transport 
calculation. To this purpose we shall use the numerical renormalization group
(NRG) in a rather unconventional way: We shall extract the $T=0$ scattering 
phase shifts directly from the NRG spectrum\cite{ALPC,Borda} and use these 
to compute the conductance by applying the Landauer-B{\" u}ttiker formula. 
Our calculations  are in good agreement with experiments and 
 reproduce the results in special limits as obtained in  
[\onlinecite{Hofstetter02}] and [\onlinecite{real}].

\section{The model}
\label{sec:model}

\begin{figure}[hbt]
\includegraphics[width= 2.3in]{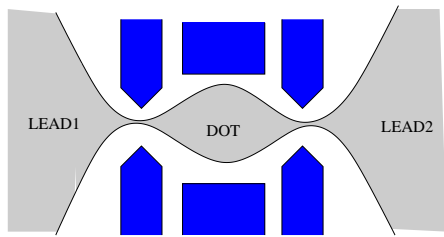}
\caption{Lateral quantum dot geometry. The dot is formed by 
depletion of the two-dimensional electron gas underneath the gates.
Dark areas denote the positions of the gate electrodes, while 
the black lines indicate the border between  depleted (white) 
and undepleted (grey) regions of the 2DEG.
\label{lateral_dot}
}
\end{figure}

In this paper we focus our attention to the lateral quantum dot 
system shown in  Fig.~\ref{lateral_dot}. Such a quantum dot is usually 
formed by gate depletion in a two-dimensional electron structure. 
The conductances  between the dot and the leads can be controlled 
by applying a  voltage on the gates separating the dot region from the 
leads. In this system Coulomb blockade develops when the dimensionless 
conductances $g_{L,R} \equiv G_{L,R}/(2e^2/h)$ between the dot 
and the left and right leads drop below 1: In this regime, electrons 
have to pay typically an energy of the order of the charging energy 
$\sim E_C$ to get onto the dot, and therefore transport through the structure is suppressed 
when the temperature drops below $E_C$. 
Since Coulomb blockade develops when the lowest propagating mode 
in the point contacts (between dot and leads) is at pinch--off, 
all higher modes can be neglected. 
The leads can thus be modeled by a single conduction electron mode  
each\cite{MF}. Note that this is specific to lateral dot systems: 
In vertical dots (see e.g.~[\onlinecite{Sasaki00,PG1}]), many conducting 
modes participate in transport. 

The Kondo effect may appear at temperatures  $T \ll E_C$ if the 
 ground state of the isolated dot is degenerate. In this case quantum fluctuations 
to the leads are relevant, and may lead to the formation of a Kondo state 
at an energy scale $T_K$. This condition is obviously fulfilled if there is an odd number of 
electrons on the dot, since then the ground state of the dot has a Kramers 
degeneracy. The Kondo effect then emerges if we lower the temperature 
below the Kondo temperature\cite{Glazman_JLTP}  
\begin{equation} 
T_K \sim \delta E \; \exp\{{-E_C / (g_L + g_R) \delta E}\} \; 
\label{T_K}
\end{equation} 
where $\delta E$ is the mean single--particle level spacing in the dot.

However, Kondo effect can also be observed in dots with an even number of 
electrons.  To understand the physics that leads to the Kondo effect in this 
case, let us first study an isolated quantum dot. 
A generic model of an isolated quantum dot can be written as\cite{ABG}
\begin{equation} 
H_{\rm dot} = \sum_{ns} \epsilon_{n} d^\dagger_{ns} d^\pdag_{ns}
+ E_C \left(\hat N - N_0 \right)^2 - E_S\hat{\bf S}^2 - B{\hat S}^z, 
\label{H_dot}
\end{equation} 
where $\hat N=\sum_{ns} d^\dagger_{ns} d^\pdag_{ns}$ is the total number of 
electrons on the dot, and 
$\hat {\bf S} = \sum_{nss'} d^\dagger_{ns} 
\left(\hat{\bm{\sigma}}_{ss'}/2\right) 
d^\pdag_{ns'}$
is the total electronic spin on the dot.  
The operators $d^\dagger_{ns}$ create electrons on a single particle level 
of the dot, labelled by the spin $s$ and a discrete quantum number $n$.
Eq.~(\ref{H_dot}) describes the electron-electron interaction
at the  mean field level. In general, more complicated interaction 
terms should  be present in $H_{\rm  dot}$\cite{large_N}. These terms 
are, however, relatively 
small for dots with a large number of electrons and furthermore, they do not 
influence our discussion of the singlet triplet transition below, therefore we 
shall neglect them.

\begin{figure}[htb]
\includegraphics[width= 3in]{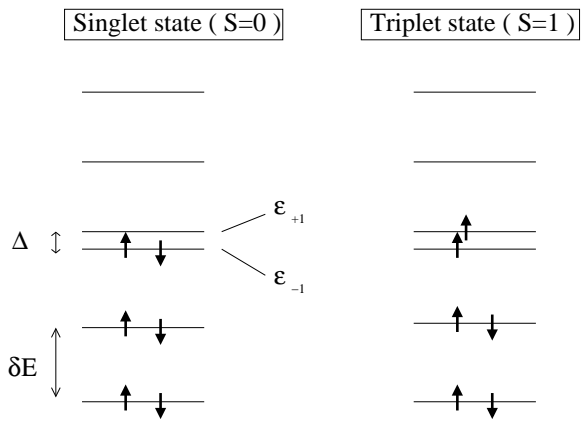}
\caption{Possible triplet and singlet ground states of an isolated dot described by 
Eq.~(\ref{H_dot}). 
\label{fig:gs}
}
\end{figure}

The parameter $N_0$ in Eq.~(\ref{H_dot}) denotes the dimensionless 
gate voltage, 
and it sets the average number $\langle \hat N \rangle$ of electrons on the dot, 
$B$ is the Zeeman field, and 
$E_C$ and $E_S$ stand for the charging energy and the Hund's rule coupling,
 respectively. 
In the rest of the paper we shall focus our attention to 
the case where $N_0$ is close to an even integer, and we shall assume
$E_C \gg \delta E > E_S$, characteristic of lateral dots with a large number of electrons\cite{ABG}.  
Under these conditions,  the ground state of Eq.~(\ref{H_dot}) 
is typically a singlet state shown on the left in 
Fig.~\ref{fig:gs}, and therefore no Kondo effect occurs.  
If, however,  the  last occupied ($\epsilon_{-1}$) 
and the first empty state ($\epsilon_{+1}$) happen to be close enough to each other, so 
that $\Delta \equiv \epsilon_{+1}-\epsilon_{-1} < 2 E_S$, then the system will form a triplet state
to gain energy from the Hund's rule coupling by rearranging 
the level occupancy (see Fig.~\ref{fig:gs}). In this case the ground state
is threefold degenerate and a Kondo state  can be formed. This transition occurs when the 
energy difference between  the singlet and triplet states approximately vanishes:
\[
\Delta_{\rm ST} \equiv \epsilon_{+1}-\epsilon_{-1} - 2 E_S \approx 0\;.
\]

The Kondo effect is driven by virtual charge fluctuations to the leads.
Since the  coupling between the leads and the dot is weak, 
$g_L,\;g_R \ll 1$,  we shall  describe these fluctuations in terms of the following 
simple tunneling Hamiltonian: 
\begin{eqnarray} 
H_{\rm leads} &=& 
\sum_{\alpha k s}\xi^\pdag_k a^\dagger_{\alpha k s} a^\pdag_{\alpha k s}, 
\quad \alpha=R,L
\label{H_leads} \\
H_{\rm tunneling} &=& \sum_{\alpha n k s} t^\pdag_{\alpha n} 
a^\dagger_{\alpha k s} d^\pdag_{ns} 
+ {\rm H.c.}
\label{H_tunneling}
\end{eqnarray} 
The operators $a^\dagger_{\alpha k s}$ in Eq.~(\ref{H_leads}) 
create electrons with momentum 
$k$, energy $\xi_k$  and spin $s$ in the left and right leads ($\alpha=L,R$), 
respectively. Eq.~(\ref{H_tunneling}) describes the tunneling between the dot 
states (labelled by the spin $s$ and a discrete quantum number $n$) and 
the leads.

In the vicinity of the singlet-triplet transition charge (and spin) fluctuations on the dot
are dominated by fluctuations to the states $n=\pm1$. It is therefore sufficient to restrict 
the summations in Eqs.~(\ref{H_dot}) and (\ref{H_tunneling}) and keep only these 
two states:
\begin{equation}
\epsilon_n = n\;\Delta/2, \quad n=\pm 1.
\label{TLS}
\end{equation}
This projection will not crucially influence our results, and the two-level model
is expected to  capture all {\em universal} 
aspects of the problem  until the relevant energy scales of the system 
that govern the singlet-triplet transition ({\em i.e.}, the Kondo scales, 
$B$, and $\Delta_{\rm ST}$) 
are all small compared to all other energy scales ($E_C$, $\delta E$ etc.) in 
the problem.
However, when projecting to these two states, one also has to keep in mind that 
at energy scales larger than $\max\{ \delta E, E_C \}$ fluctuations the the 
{\em excited} states of the dot destroy those coherent contributions that give 
rise the Kondo effect. 
Therefore, at the same time the effective band-width in Eq.~(\ref{H_tunneling})  
has to be reduced to $D \equiv \max\{ \delta E, E_C \}$\cite{cut-off}.

In the following sections we shall study the  conductance of the dot when one 
gradually drives the system through the triplet singlet transition. 
There are several experimental techniques to achieve this goal and 
tune $\Delta$ and thus $\Delta_{\rm ST}$. 
The easiest method is to apply a magnetic field $H_\perp$  perpendicular to the 
plane of the dot.  The primary effect of $H_{\perp}$ is to   change the orbital wave functions
and thus to control  the value of  $\Delta(H_\perp)$.\cite{Sasaki00,vanderWiel02} 
By linearizing $\Delta(H_\perp)$ in the vicinity of the transition, 
one can thus attempt to make a direct 
comparison of the experimental data with the (calculated) linear conductance across 
the system $G = (2e^2/h)g(\Delta)$. 
Unfortunately, $H_\perp$ also induces a Zeeman field $B(H_\perp)$: 
While this Zeeman splitting $B$  remains small compared to the 
orbital effect  because of the smallness of the electron 
effective mass and the $g$-factor in GaAs\cite{induced_review}, yet it is 
comparable to the experimentally observed Kondo scale on the triplet 
side\cite{vanderWiel02}, and leads to some complications when trying to make 
 a direct comparison with the experimental data.
The difficulties related to Zeeman splitting were avoided in the 
experiment [\onlinecite{Kogan02}] by tuning the singlet--triplet 
transition via the Stark effect due to an inhomogeneous gate 
potential.  

Another way to avoid this problem is to use {\em triangular dots} where the degeneracy of the states
$\epsilon_{\pm 1}$ can be directly controlled by changing the shape of the 
dot\cite{David}. To investigate the 
dependence of $g(\Delta, B)$ on $B$ experimentally, one 
may in addition apply a strong in-plane field $H_{\parallel}$, which 
only generates a Zeeman splitting without influencing $\Delta$ essentially.

With the method outlined in Section~\ref{sec:FS} we cannot compute the finite 
temperature conductance of the dot. 
However, the temperature ($T$), source--drain bias ($V$), and
Zeeman field ($B$) dependences of the conductance are expected to be
qualitatively (but not necessarily  quantitatively) similar to each 
other. Therefore many of our results can be used to understand qualitatively the
finite  temperature and finite bias behavior of the dot in the absence of 
the Zeeman field $B$ by replacing $B$ with $V$ or $T$.

\section{Linear conductance}

As discussed above, in the vicinity of the singlet-triplet transition the 
Hamiltonian  of the dot can be truncated to that of a two-level 
system (Eqs.~(\ref{H_dot}) and (\ref{TLS})). 
The tunneling, Eq.~(\ref{H_tunneling}), 
couples the two levels to the two leads, and the four tunneling 
amplitudes $t_{\alpha, n}$ form a $2\times 2$-matrix
\[
\hat t = \left(
\begin{array}{cc}
t_{L,+1} &  t_{R,+1} \\
t_{L,-1}  &  t_{R,-1} 
\end{array}
\right).
\]

Consider first the {\it special case} when one of the eigenvalues of the matrix $\hat t$ 
is zero, while another one is finite\cite{Hofstetter02}. Obviously, in this case the 
dot effectively interacts only with a {\it single} species (a single ``channel'') of conduction 
electrons. When the dot is in the triplet state, a single electronic channel can screen 
only half of its spin\cite{NB}. Accordingly, the system should exhibit a quantum phase 
transition\cite{Vojta02}: the ground state changes its symmetry from a singlet to a 
doublet as $\Delta$ decreases below a certain critical value $\Delta_C\sim E_S$. 
The linear conductance $g(\Delta,0)$ across the system at $T=0$ and $B=0$ is then 
strongly $\Delta$-dependent\cite{Hofstetter02}: 
$g(\Delta,0) \propto \theta(\Delta_C -\Delta)$. 
At $\Delta < \Delta_C$ (when the dot is in the triplet state) the conductance is a 
monotonically decreasing function of $B$. To the contrary, at $\Delta > \Delta_C$ 
the conductance as a function of $B$ first increases, and then drops 
with the increase  of $B$. At a fixed finite $B$ the conductance 
$g(\Delta,B)$ is expected to be a 
smooth non-monotonic function of $\Delta$ with a broad asymmetric 
peak near the  transition. 

\begin{figure}[bh]
\includegraphics[width= 0.9\linewidth]{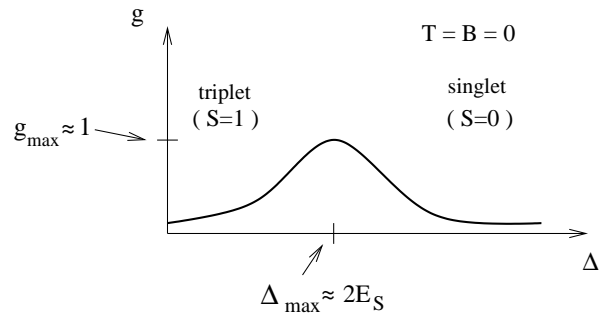}
\caption{
\label{fig:sketch}
Sketch of the zero temperature conductance as a function of the splitting 
between the two dot levels for generic tunneling amplitudes. 
}
\end{figure}

In a {\it general situation}, however, $\det \hat t \neq 0$, and both eigenvalues of the 
matrix $\hat t$ are finite. In this case, the dot is coupled to {\it two} electronic channels, 
which is sufficient in order to completely screen the dot's spin\cite{NB} even in 
its triplet state. As a result, the ground state of the system is a {\it singlet} at all values of 
$\Delta$\cite{PG1,PG2,Izumida2}. Accordingly, when the dot is coupled to the leads, 
the singlet-triplet transition turns to a crossover. As we discuss below, 
the conductance $g(\Delta,0)$ 
at $T=0$ and $B=0$ is then expected to be a smooth non-monotonic function 
of $\Delta$, slowly (logarithmically) decreasing\cite{PG2} at large $|\Delta-2E_S|$
(see Fig.~\ref{fig:sketch}). Far 
away from the crossover, the conductance is a non-monotonic function of 
$B$: it first raises and may reach a value $g\sim 1$, and then drops again as $B$ increases. 
This has been shown for both the triplet side of the crossover\cite{real,Izumida1} 
($\Delta\ll E_S$), and for the singlet ($\Delta\gg E_S$) side of it\cite{PG2,Zeeman}. For a  fixed small $B$, however, 
the conductance $g(\Delta,B)$ can be  an even more complicated  function 
of $\Delta$ with  a local minimum at some $\Delta <  E_S$ and maxima at the triplet side of the transition ($\Delta \ll  E_S$) and  around $\Delta\sim E_S$.  

In this paper, we focus on a particular limit when both eigenvalues of $\hat t$ are finite, 
but still very different: $\left|\det \hat t\right| \ll \left({\rm Tr}~\hat t\right)^2$. 
In addition, we constrain our considerations to a special subset\cite{Izumida2,Izumida1} 
of the matrices $\hat t$ that can be parametrized as
\begin{equation}
\hat t = \frac{1}{\sqrt 2}\left(
\begin{array}{cc}
v_{+1} &  v_{+1} \\
v_{-1}  &  - v_{-1} 
\end{array}
\right),
\quad
v_{\pm 1}\neq 0.
\label{amplitudes}
\end{equation}
%G
This tunneling matrix arises, {\em e.g.},  for a fully symmetric arrangement when the states 
$n=\pm1$ are symmetric/antisymmetric under reflection. 
The choice (\ref{amplitudes}), although by no means general, nevertheless captures 
the essential physics of the system. 
At the same time, as we demonstrate shortly below,
the choice (\ref{amplitudes}) allows one to express the conductance $g(\Delta, B)$ 
at $T=0$ in terms of simple phase shifts that can be extracted from the 
finite size NRG spectra\cite{Borda},  
or in some cases can be related to thermodynamic quantities\cite{Izumida2,Izumida1}. 
%[??????????CAN'T WE DO THE PHASE SHIFT CALCULATION FOR A GENERAL t MATRIX ALSO?????????]
%G  I am not sure what your question is, so maybe my  answer does not answer your question. 
% We can, of course, determine the phase shifts for any geometry. The problem is that, in general, 
% the angle theta in the transformation depends on Delta, B etc, and this influences the amplitude of the 
% conductance. 

Since the ground state of the system is not degenerate, electrons scatter 
elastically at $T=0$. The scattering amplitudes $S_{s;\alpha,\alpha'}$ of 
an electron with spin $s$ from lead $\alpha '$ to lead $\alpha$ form the 
scattering  matrix ${ S}_{s\alpha;s'\alpha'} 
= \delta_{ss'}\; S_{s;\alpha,\alpha'}$.  
The $4\times 4$ unitary 
%G corrected 
%  $2\times 2$ -> $4\times 4$
matrix ${\hat S}$ can be diagonalized by 
a rotation in the $R-L$ space to the new basis of {\it channels} $n=+1$ and $n=-1$,
\begin{equation}
U\hat S U^\dagger = {\rm diag}\left\{e^{2i\delta_{ns}}\right\}, 
\quad
U = e^{i\vartheta \tau^y} e^{i \varphi  \tau^z}. 
\label{diag}
\end{equation}
Here $\tau^i$ are the Pauli matrices acting in the $R-L$ space. 
The linear conductance at $T=0$ is related to the off-diagonal 
elements of $\hat S$ by the Landauer formula,
\[
g = \frac{1}{2}\sum_s \left|S_{s;RL}^2\right|,
\]
and, by making use of Eqs.~(\ref{diag}), can now be expressed via the scattering 
phase shifts $\delta_{ns}$ at the Fermi energy.
%:
%\begin{equation}
%g = \sin^2(2\vartheta)\frac{1}{2}\sum_s\sin^2\left(\delta_{+1,s} - \delta_{-1,s}\right) .
%\label{formula}
%\end{equation}
In general, the angle $\vartheta$ and thus $g$  depend explicitly on the parameters of the 
microscopic Hamiltonian.  However, with the choice Eq.~(\ref{amplitudes})
the tunneling part of the Hamiltonian can be trivially diagonalized with the introduction of even and 
odd states (corresponding to $\vartheta = \pi/4$, $\varphi =0$)
\[
\left(\begin{array}{c}
c_{+1, k s} \\
c_{-1, k s} 
\end{array}\right)
= 
{1\over \sqrt{2}}
\left(\begin{array}{c}
a_{Rks} + a_{Lks} \\
a_{Rks} - a_{Lks}
\end{array}\right),
\]
and the Hamiltonian takes the form
\begin{equation}
H = \sum_{n k s} 
\left[ 
\xi^\pdag_k c^\dagger_{nks} c^\pdag_{nks} 
+  v^\pdag_{n}  \left(c^\dagger_{nks} d^\pdag_{ns} 
+ {\rm H.c.}\right)
\right]
+ H_{\rm dot}\;.
\label{H_new}
\end{equation}
The Hamiltonian (\ref{H_new}) with $H_{\rm dot}$ 
given by Eqs.~(\ref{H_dot}),(\ref{TLS}) obviously 
conserves the total number of fermions in 
each channel, ${\cal N}_n = \sum_{ks} \left[c^\dagger_{nks} c^\pdag_{nks} 
+ d^\dagger_{ns} d^\pdag_{ns}\right]$, implying that the scattering matrix 
is diagonal in $n$.

In this case the conductance simply reads \cite{Georges,real}:
\begin{equation}
g = \frac{1}{2}\sum_s\sin^2\left(\delta_{+1,s} - \delta_{-1,s}\right) .
\label{conductance}
\end{equation}

Following Nozi\`eres, at small $B$ the phase shifts can be expanded 
to linear order\cite{Nozieres,Glazmanfirst}. As explained in [\onlinecite{Glazmanfirst}],
a simple analysis shows that in the presence of a Zeeman field the conductance at low fields 
behaves as 
\begin{equation}
g(\Delta,B) \approx g(\Delta,0) 
+ \left[1-2g(\Delta,0)\right]\left(\frac{B}{B^*(\Delta)}   \right)^2
\label{small_B}
\end{equation}
where the scale $B^*$ is a $\Delta$-dependent energy scale and  
\[
g(\Delta,0) = \sin^2\left[\delta_{+1}(\Delta)-\delta_{-1}(\Delta)\right].
\]
In other words, far away from the transition the conductance should increase 
with $B$ for $g(\Delta,0)<g^* \approx 1/2$ 
while it should typically decrease in the transition region, 
where  $g(\Delta,0)>g^*$. 
We have to emphasize that the value $g^* \approx 1/2$, see Eq.~(\ref{small_B}),  
is specific to the choice (\ref{amplitudes}), and not universal.

%G modified this paragraph since the procedure described did not work...

In the following, we shall compute the phase shifts using the powerful 
numerical renormalization group and determine the conductance 
using  Eq.~(\ref{conductance}).

\section{Numerical RG, finite size spectra, and phase shifts} 
\label{sec:FS}

The \emph{Numerical Renormalization Group} (NRG)
was originally developed 
by Wilson to solve the Kondo problem\cite{Wilson75}. 
In contrast to scaling approaches, %\cite{Anderson70}, 
it is non--perturbative and does not encounter any 
logarithmic singularities at low energy. 
Since then it has been extended to the calculation of dynamical quantities
\cite{Costi94,Hofstetter00} and has been applied to a variety of 
quantum--impurity problems,  including transport calculations for interacting 
quantum dots \cite{Hofstetter01,Borda}. 

In principle, the NRG method enables us to compute the expectation value of 
any local operator. Thus one plausible method to obtain the phase shifts would be 
to compute the occupation numbers $N_{ns} = \left\langle d^\dagger_{ns} d^\pdag_{ns} 
\right\rangle$ in the ground state by NRG and then  use the Friedel sum rule
to get the phase shifts as \cite{Hewson,Langreth}:
\begin{equation}
\delta_{ns} = \pi N_{ns}.
\label{eq:Friedel}
\end{equation}
 Unfortunately, this procedure  turns out to deliver  unphysical results, especially 
on the triplet side. The origin of this problem is probably related to the
relatively small bandwidth in our calculations and the large logarithmic 
corrections that appear on the triplet side: As is obvious from the 
derivation of the Friedel sum rule\cite{Hewson},  
there is, in general, a {\em correction term} 
to Eq.~(\ref{eq:Friedel}) which vanishes in the infinite band width limit. 
In our calculations, the Kondo temperature is comparable to the 
bandwidth $D$. Furthermore, on the triplet side of the transition 
the convergence to the ground state is rather slow due to the
large ferromagnetic residual exchange coupling, generic to 
underscreened Kondo models, that  can even lead to singular 
non-Fermi liquid  properties\cite{Piers}.  
Therefore we cannot use the Friedel sum rule to determine the 
conductivity of the dot. Nevertheless, the Friedel sum rule should 
definitely work in the limit of very large cut-offs. It therefore gives us a 
{\em useful tool} to understand qualitatively the evolution of the phase shifts 
and the conductance.

Fortunately, as discussed below, we can determine the phase 
shifts {\em directly}  from the NRG spectra without making use of the 
Friedel sum rule\cite{ALPC}. This method has been first used 
to compute the conductance of a mesoscopic double dot system\cite{Borda}. 
To do this, we only have to know that the model flows 
to a universal {\em Fermi liquid}  fixed point that can be characterized by 
four phase shifts. 

To explain the method, let us briefly discuss Wilson's NRG procedure.
The key idea of Wilson is to map the original problem
to a semi--infinite chain.  After this transformation our 
two--channel Hamiltonian 
 maps onto a system of two parallel tight--binding chains 
coupled to the dot 
%(see Fig. \ref{fig:double_chain}).
\begin{eqnarray}
H & = &H_{\rm dot} + \sum_{n, s} {\tilde v}_{n} 
(d_{ns}^\dagger c^\pdag_{0, ns} + {\rm h.c.}) 
\nonumber \\
%H_{band} = 
& + & \sum_{ n, s}\sum_{i=0}^\infty 
\xi_i \left(c^\dagger_{i, n s} c^\pdag_{i+1, n s} + h.c.\right) \;,
\label{eq:wilson}
\end{eqnarray}
%
%\begin{equation}
%H_{band} = \sum_{n=0 \atop i, \sigma}^\infty 
%\xi_n \left(c^\dagger_{n i \sigma} c_{n+1\; i \sigma} + h.c.\right)
%\label{eq:wilson}
%\end{equation}
%
where the hopping matrix elements  decay exponentially, 
$\xi_i \approx \frac{1 + \Lambda^{-1}}{2} \Lambda^{-i/2}$, with $\Lambda$ 
the NRG discretization parameter.  
In our calculation we take $\Lambda=3$.  
The $c_{i,ns}$ represent conduction electron 
excitations at a  length scale $\sim \Lambda^{i/2}$. 
The NRG assumes a flat dispersionless 
density of states for the electrons and all energies 
are measured in units of the bandwidth $D$.
The operator $c_{0, n s}$ is simply the local field operator 
defined as  $c_{0, n s} \equiv {1\over\sqrt{2}} \int_k dk 
c_{ k n s}$,  and it is only this operator that tunnel--couples to 
the interacting impurity part, $H_{\rm dot}$. 
Note also that the discretization procedure somewhat renormalizes 
the original parameters of the Hamiltonian, 
%${v}_{n}\to{\tilde v}_{n}$, 
which are therefore not identical to the  ones in Eq.~(\ref{amplitudes}).

Eq.~\ref{eq:wilson} can be rewritten in a more inspiring way as 
\begin{eqnarray}
{\tilde H}_{N+1} &= &\Lambda^{1/2}  \; {\tilde H}_N + \sum_{n,s} 
\left(c^\dagger_{N n s} c^\pdag_{N+1\; n s} + h.c.\right)  \\
H & = & \lim_{N\to\infty} {1 + \Lambda^{-1} \over 2} \Lambda^{-(N-1)/2} {\tilde H}_N\;,
\label{eq:iterative}
\end{eqnarray}
where we have introduced the rescaled Hamiltonians  
\begin{eqnarray}
{\tilde H}_{N} & \equiv &{2\over 1 + \Lambda^{-1}}\Lambda^{(N-1)/2} H_N\;, \\
H_N & \equiv &H_{\rm dot} + \sum_{n, s} { \tilde{v}}_{n} 
(d_{ns}^\dagger c^\pdag_{0, ns} + {\rm h.c.})  \nonumber \\
& + &  \sum_{ n, s}\sum_{i=0}^{N-1} 
\xi_i \left(c^\dagger_{i, n s} c^\pdag_{i+1, n s} + h.c.\right) \;,
\end{eqnarray}
%
%[*********I'VE CHANGED THE DEFINITION OF THE RESCALED HAMILTONIAN $H_N$ 
%(FOLLOWING KRISHNAMURTHY ET AL.), PLEASE CHECK WHETHER THIS IS CONSISTENT 
%WITH YOUR CFT CALCULATION*************]
and $H_{0}$ denotes the impurity Hamiltonian,
$H_{\rm dot}$, tunnel-coupled to $c_{0,p s}$. 
Eq.~(\ref{eq:iterative}) can  be solved by iterative 
diagonalization, by keeping in each step only the lowest, most 
relevant levels. In our calculation, conservation  of total charge 
and the $z$--component of the 
total spin
\begin{eqnarray}
Q &= & \sum_{ n, s}\sum_{i=0}^\infty 
(c^\dagger_{i, n s} c_{i, n s}-1/2) \\
S_z &= &\sum_{ n, s}\sum_{i=0}^\infty 
s \; c^\dagger_{i, n s} c^\pdag_{i, n s}
\end{eqnarray}
have been exploited to increase numerical  efficiency.

\begin{figure}[hb]
\includegraphics[width=0.95\linewidth]{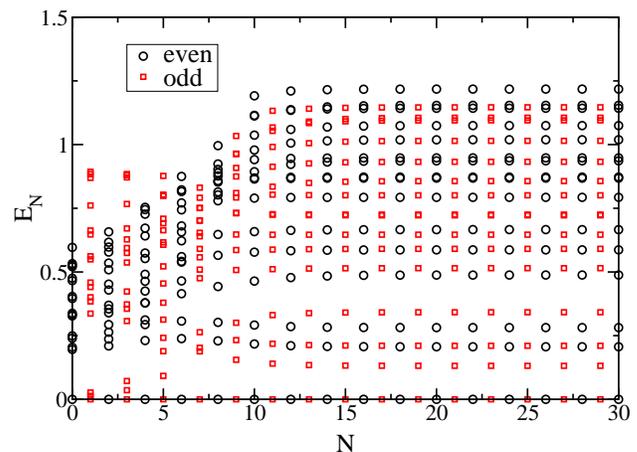}
\caption{\label{fig:NRG_flow} 
A typical NRG level flow on the triplet side for \mbox{$E_C=1$}, $E_S=0.25$, $\Delta = 0.2$, 
$v_{+1}=0.67$, $v_{-1}=0.37$ and $B=0.01$. 
The results were obtained with $\Lambda=3$ and keeping $N=1000$ levels.
}
\end{figure} 

A typical  example of the spectrum of ${\tilde H}_N$ vs. iteration number is 
shown in Fig. \ref{fig:NRG_flow}. At large iteration numbers 
({\em i.e.}~small energy scales), the spectrum  remains invariant, indicating the 
presence of a stable field  theory (fixed point) which describes the low 
energy behavior of our model.  

To interpret the spectrum, we have to 
keep in mind that in the NRG procedure the length scale $L$ of the system 
corresponds to $L\sim \Lambda^{N/2}$, and therefore, apart from some 
multiplicative factor, the spectrum of  ${\tilde H}_N$ is the spectrum 
of a system of size $L$ in units of $2\pi v_F / L$. 
This fixed point NRG spectrum is {\em universal} 
and can be characterized by boundary conformal field theories\cite{boundaryCFT}.  
In our case the fixed point spectrum is that of a Fermi liquid, and  
is therefore identical to that of the following simple model
\begin{equation}
H_{\rm fp}  =  \sum_{n,s} \int_{-L/2}^{L/2} dx\;
\psi^{\dagger}_{ns}(x) (-i \partial_x + U_{ns} \delta(x))
 \psi_{ns}(x) 
\end{equation}
where the $ \psi_{ns}(x) $'s denote chiral fermionic fields obeying 
periodic and anti-periodic boundary conditions for even and odd $N$, 
respectively. This statement means that the full many body--spectrum of our  
model is identical to that of the following simple effective Hamiltonian
(in Fourier space):
\begin{equation}
H_{\rm fp}  =  {2\pi\over L} \sum_{n,s} \sum_{q} 
(q - {\delta_{n,s} \over \pi} )
\psi^\dagger_{q,n  s} \psi^\pdag_{q,n  s}\;,
\label{eq:eff_spectrum}
\end{equation}
where the phase shifts $\delta_{n,s}$ are related to 
the couplings $U_{n s}$ and $q$ takes integer (half-integer) 
values for even (odd) iterations. In more precise terms:
\begin{equation}
\lim_{N\to \infty} {\tilde  H}_{N} = C \;{L \over 2\pi} H_{\rm fp}\;, 
\end{equation}
where $C$ is a $\Lambda$--dependent constant to be determined numerically.
The quantum numbers $Q$ and $S_z$ in the fixed point theory are defined as 
\begin{eqnarray}
Q^{\rm fp} &= & \sum_{q, n, s}
:\psi^\dagger_{q, n s} \psi_{q, n s}:\;, \\
S^{\rm fp}_z &= &\sum_{q, n, s}
s \; \psi^\dagger_{i, n s} \psi_{i, n s}\;,
\end{eqnarray}
where $:\dots:$ denotes normal ordering with respect to the ground state.

To illustrate how well this procedure works we have enumerated in Table 
\ref{table_spectrum} the lowest eigenstates and their quantum numbers 
in the NRG spectrum together  with those of the effective Hamiltonian 
$ H_{\rm fp}$: All quantum numbers are in agreement and the eigenvalues 
match up to four digits. In our calculations we have not kept track of the 
conserved quantum numbers $N_n$. Therefore,  while we could tell the values 
of the two spin up and spin down phase shifts, $\delta_s$ and ${\tilde \delta}_s$, in some cases we were 
unable to tell to which channel they belonged. Since, however, 
Eq.~(\ref{conductance}) is symmetrical under the exchange of 
$n=\pm$,  this caused no problem in the calculation of the conductance. 

This method enables us to extract the phase shifts {\em directly from 
the finite size spectrum}. The great advantage of this procedure is that 
it  works directly on the fixed point Hamiltonian and therefore eliminates 
the  problem of high energy degrees of freedom  that occurred when 
using the Friedel sum rule. This method can be applied to any problem 
where the ground state is that of a Fermi liquid.

%% The tables I and III in the text are so big that they
%  do not fit on one page when printed in preprint format.
%  They are only correctly formatted in two-column format,
%  for which, instead of the \documentstyle command in the first line,
%  the alternative command 
%
%\documentstyle[aps,prb,amsfonts,epsf]{revtex}
%
% in the preamble of this .tex file should be activated.
%
%\widetext
\begin{table}
\begin{tabular}{c|cccc}
\hline\hline
\phantom{n}state \phantom{n} &   \phantom{n} ${\tilde E}_{\rm NRG}/C$ \phantom{n} &   \phantom{n}       $L\;E_{\rm fp}/2\pi$  \phantom{n} &      \phantom{n}$Q$  \phantom{n}    &  \phantom{n}$S_z$  \phantom{n}\\
\hline
1       &       0.1288  &       $1-{\delta_\uparrow\over\pi} =0.1288$   & 1 & 1/2 \\
2       &       0.1757  &       ${\delta_\downarrow\over\pi} =0.1757$   & -1 & 1/2 \\
3       &       0.3045  &       $1-{\delta_\uparrow\over\pi}+ {\delta_\downarrow\over \pi} =0.3045$     & 0 & 1 \\
4       &       0.3661  &       ${{\tilde \delta}_\downarrow\over\pi} =0.3661$  & -1 & 1/2 \\
5       &       0.4157  &       $1-{{\tilde \delta}_\uparrow\over\pi} =0.4157$  & 1 & 1/2 \\    
6       &       0.4949  &       $1+{{\tilde \delta}_\downarrow\over\pi}-{{\delta}_\uparrow\over\pi} =0.4949$ & 0 & 1 \\
7       &       0.5418  &       ${{\tilde \delta}_\downarrow\over\pi}+{\delta_\downarrow \over \pi} =0.5418$& -2 & 1 \\
8       &       0.5445  &       $2-{{\tilde \delta}_\uparrow\over\pi} - {{ \delta}_\uparrow\over\pi}  =0.5445$  & 2 & 1 \\
9       &       0.5843  &       ${{\tilde \delta}_\uparrow\over\pi} =0.5843$    & -1 & -1/2 \\
10      &       0.5914  &       $1-{{\tilde \delta}_\uparrow\over\pi} + {\delta_\downarrow\over\pi} = 0.5914$  & 0 & 1 \\
\hline\hline
\end{tabular}\vspace*{4mm}
\caption[Table-1]{
\label{table_spectrum}  Comparison of the even iteration 
fixed point NRG spectrum in Fig.~\ref{fig:NRG_flow} and that of the effective 
field theory, Eq.~\ref{eq:eff_spectrum}. We used $C=1.6023$, and the phase shifts 
were ${\delta_\downarrow\over\pi} =0.1757$, 
${{\tilde \delta}_\downarrow\over\pi} =0.3661$, 
${{\tilde \delta}_\uparrow\over\pi} =0.5843$, and ${\delta_\uparrow\over\pi} =0.8712$.
$H_{\rm fp}$ reproduces all the quantum numbers correctly and the NRG spectrum with 
a four digit precision.
}
\end{table}

\section{Results}

\begin{figure}[h]
\begin{center}
\includegraphics[width=0.75\linewidth]{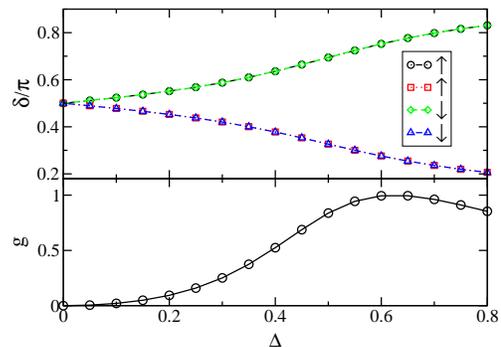}
\end{center}
\caption{
\label{fig:phase_shift+cond_Delta}
Phase shifts (top) and corresponding conductance (bottom) as a function of 
orbital splitting $\Delta$ in the absence of the Zeeman field at temperature $T=0$. 
Parameters are chosen as $E_C = 1$, $E_S = 0.25$, $v_{+1}=0.67$ and $v_{-1}=0.50$. }
\end{figure}

In Fig.~\ref{fig:phase_shift+cond_Delta} we have plotted the typical splitting dependence 
of the phase shifts for $B=0$ and the corresponding conductance. 
The qualitative behavior of the phase shifts can be 
understood in terms of the Friedel sum rule Eq.~(\ref{eq:Friedel}) as follows:
On the triplet side 
of the transition the spin of the dot is completely screened by the electrons in the leads,
which corresponds to a phase shift $\pi/2$ in both channels. On the singlet side, however, 
both dot electrons occupy the level $n=-1$. Therefore the phase shift in channel $n=-1$ should 
approach $\delta_{-1,s}\approx \pi$ on this side, while the phase shift $\delta_{+1,s}$
must go to zero. By Eq.~(\ref{conductance}) this implies that the conductance must approach 
0 on both sides of the transition while it has a maximum $g_{\rm max}\sim 1$ around 
$\delta_{-1,s}-\delta_{1,s}\approx \pi/2$. 
Note that the transition point is shifted with respect to the bare value 
$\Delta=2 E_S=0.5$ for the isolated dot due to correlation effects.

\begin{figure}[h]
\begin{center}
\includegraphics[width=0.8\linewidth]{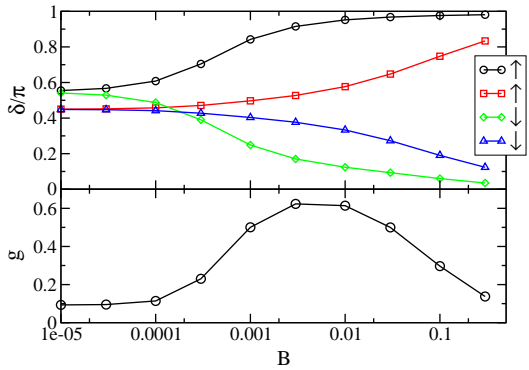}
\end{center}
\caption{
\label{fig:phase_shift+cond_B}
Phase shifts (top) and corresponding conductance (bottom) as a function of Zeeman field $B$ 
on the triplet side of the transition at zero temperature. Parameters are 
$E_C = 1$, $E_S = 0.25$, $\Delta=0.4$, $v_{+1}=0.67$ and $v_{-1}=0.25$.
}
\end{figure}

Application of a Zeeman field $B$ has a more complex effect, which to some degree 
has been discussed in [\onlinecite{real}]. On the triplet side,
$\Delta \ll E_s$ we have two exponentially separated  Kondo scales, $T_{K}^> \gg T_{K}^<$.
For Zeeman fields $B \ll T_{K}^>,T_{K}^<$ all four phase shifts approximately 
equal $\pi/2$, corresponding to a reduced conductance. However, for intermediate fields 
$T_{K}^< \ll B \ll T_{K}^>$, the Kondo effect in channel 
$n=-1$ is suppressed, and therefore the phase shifts in this channel are expected 
to be around $\delta_{n=-1,\uparrow}\approx \pi$ and 
$\delta_{n=-1,\downarrow} \approx 0$, while 
$\delta_{n=+1,\uparrow}\approx \delta_{n=+,\downarrow} 
\approx \pi/2$, leading to a conductance $g\sim 1$. Finally, for even larger fields, 
$B \gg T_{K}^<,T_{K}^>$, both Kondo effects are suppressed, the phase shifts become
 $\delta_{n=-1,\downarrow}\approx  \delta_{n=1,\downarrow} \approx 0$ and 
$\delta_{n=-1,\uparrow}\approx  \delta_{n=1,\uparrow} \approx \pi$, and the conductance
decreases again to 0. This behavior is shown in Fig.~\ref{fig:phase_shift+cond_B}.

\begin{figure}[h]
\begin{center}
\includegraphics[width=0.8\linewidth]{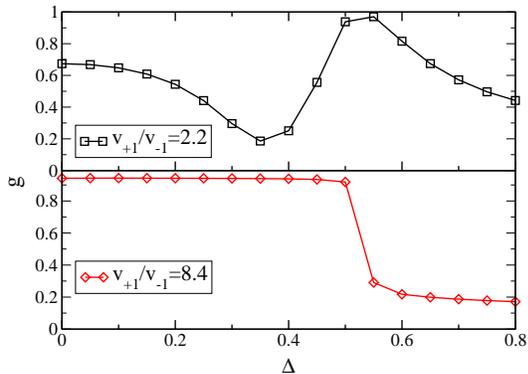}
\end{center}
\caption{
\label{fig:g(Delta)_B_fixed}
Conductance as a function of $\Delta$ for a fixed value $B=10^{-3}$ of the Zeeman field 
and coupling asymmetry ${ v}_{+1}/{ v}_{-1}= 2.2$ (top) and 
${v}_{+1}/{v}_{-1}= 8.4 $ (bottom). 
The other parameters are $E_C = 1$, $E_S = 0.25$, $v_{+1}=0.67$.}
\end{figure}

As we already mentioned in Section~\ref{sec:model}, in the experiment 
[\onlinecite{vanderWiel02}] the Zeeman field was non--negligible. It is therefore 
instructive to study the splitting--dependence of the conductance in the 
presence of a {\em finite} Zeeman field $B$. This is shown in Fig.~\ref{fig:g(Delta)_B_fixed}.
In a finite magnetic field $g(\Delta,B={\rm fixed})$ can show a very surprising 
$\Delta$ dependence for some values of $B$, and  may even have a dip on the triplet side. 
As we explain below, this dip appears because $T_K^<$ is very sensitive to the 
distance from the transition point, and therefore we may cross over from the regime 
$T_K^>, T_K^< > B$ to $T_K^> > B > T_K^<$ as we are getting farther away from the transition 
region. For intermediate  Zeeman fields $B$ 
the conductance has the shape of a smeared step function that becomes sharper as 
the ratio ${ v}_+/{v}_-$ increases and resembles  the one found in 
[\onlinecite{Hofstetter02}].

\begin{figure}[h]
\includegraphics[width=0.8\linewidth]{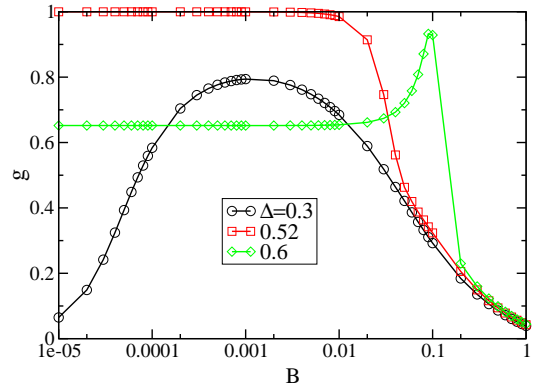}
\caption{\label{fig:conductance_vs_B_I} 
Linear conductance versus in--plane magnetic field $B$ 
for different values of the orbital splitting and the same parameters as 
in Fig.~\protect{\ref{fig:phase_shift+cond_B}}.
Notice the non--monotonic behavior (``dip'') 
on both the triplet ($\Delta = 0.3$) and singlet ($\Delta = 0.6$) sides of the 
cross-over. In the middle of the transition ($\Delta = 0.52$) the conductance 
is $g=1$ at $B=0$, and decreases monotonically as a function of magnetic field.
} 
\end{figure}

The effect of a Zeeman field on the conductance should 
qualitatively resemble that of a finite source--drain bias or temperature.
We can thus use the $B$--dependence of the conductance for fixed fixed values 
of $\Delta$ to identify the second energy scale that appears as the width of 
a dip in the zero-bias anomaly in the experiment [\onlinecite{vanderWiel02}].
The $B$--dependent conductance for three different values of $\Delta$ is shown in 
Fig.~\ref{fig:conductance_vs_B_I}. The conductance first
increases for small  Zeeman fields $B$ on both sides of the transition, but it is 
suppressed for larger values of $B$. However, it decreases monotonically in the transition 
region, where $g>g^*$ (see Eq.~\ref{small_B}). 

\begin{figure}[h]
\includegraphics[width=0.8\linewidth]{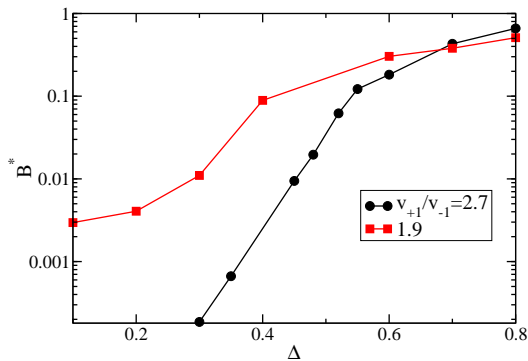}
        \caption{\label{fig:B*} 
Low--energy scale $B^*$ for $E_C =1$, $E_S=0.25$ and $v_{+1}=0.67$,  
as determined from the curvature of $g(\Delta, B)$ for small Zeeman field $B$. 
Note the saturation of $B^{*}$ on the triplet side for the larger 
$v_{-1}$. 
}
\end{figure}

We can then identify the low--energy scale $B^*$ by fixing $\Delta$ and measuring 
the curvature of  $g(B,\Delta)$. This is shown in Fig.~\ref{fig:B*}.  
On the singlet side of the transition $B^*$ increases approximately  linearly with 
$\Delta$, and practically corresponds to the energy splitting of the singlet and triplet 
multiplets. Intuitively this can be understood in the following way: Applying a Zeeman field one 
gradually pulls down the $S_z = -1$ state of the excited triplet state. This becomes 
degenerate with the singlet state if $B$ is approximately equal to the 
splitting between the triplet and the singlet states of the dot.
At this special value of 
$B$ a Kondo effect can occur between the ground state singlet and the $S_z=-1$ triplet  
state\cite{Zeeman}, giving a conductance maximum as a function of $B$. Indeed, we can 
see in Fig.~\ref{fig:conductance_vs_B_I} that on the singlet side ($\Delta = 0.6$)
the maximum conductance is indeed
approximately $g\approx 1$, indicative of the Kondo resonance. Therefore, on the singlet side
$B^*$ must be proportional to the renormalized (by quantum fluctuations) 
singlet-triplet splitting.

The origin of the dip on the triplet side is very different, as explained above.  
Far away from the transition, 
the spin $S$ of the dot is screened by two consecutive Kondo effects.  
An intermediate magnetic field suppresses the Kondo effect 
with the smaller Kondo temperature, $T_K^<$, 
and leads to a large, almost unit conductance.
Therefore, in this region, the energy  scale $B^*$ should practically correspond to the 
smallest of these two  Kondo temperatures, {\em i.e.} the one 
with smaller hybridization $v_<$. 
Far away from the triplet side, $T_K^<$ does not depend strongly on 
the presence of the singlet.
However, in the {\em vicinity} of the singlet-triplet 
transition,  quantum fluctuations to the singlet state are extremely important: 
They mix all four dot states, renormalize the value of $v_<$, 
and generate a strongly correlated state with a Kondo scale that is 
determined by the hybridization 
to the more strongly coupled  channel, $v_>$. As a consequence, in the vicinity  
of the transition these fluctuations strongly renormalize the value of $v_<$ and 
thus influence $B^*\sim 
T_K^< $. This is shown in Fig.~\ref{fig:B*}.

For the parameters of the figure the second Kondo scale would be very small in the absence of 
fluctuations to the  singlet state. Therefore, $B^*$ is  extremely sensitive to the distance from 
the triplet-singlet transition, which sets the renormalized value of $v_<$, and thus that of 
$T_K^< \sim B^*$. Far from the transition the effect of the fluctuations to the 
singlet state is small, and they 
do not renormalize $T_K^< \sim B^*$ substantially. Thus $T_K^< $  saturates on 
the triplet side and shows only a weak $\Delta$-dependence  there, 
in complete agreement with the experimental results of Ref.~[\onlinecite{vanderWiel02}]. 
However,  for larger ratios of  $v_{+1}/v_{-1}$, the  saturation value  $T_K^<(\Delta_{\rm ST}\ll0)$  becomes
 so small that it can hardly be observed. 

\begin{figure}[h]
\includegraphics[width=0.8\linewidth]{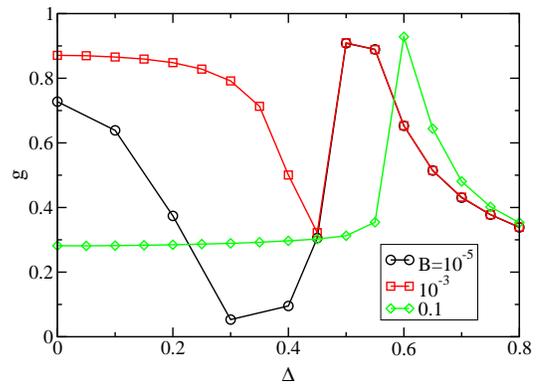}
        \caption{\label{fig:conductance_vs_splitting_I} 
        Linear conductance versus level splitting for different 
in--plane magnetic fields $B$ and a fixed ratio ${\tilde v}_{+1}/{\tilde v}_{-1}= 2.6$. 
The other parameters are the same as in Fig.~\protect{\ref{fig:phase_shift+cond_B}}.
}
\end{figure}

In Fig.~\ref{fig:conductance_vs_splitting_I} we show the $\Delta$--dependence of the conductance
for three different values of $B$. The conductance has a similar behavior for the two lower
values of $B$. In these cases, far away on the 
triplet side,  we have a situation where $T_K^< < B< T_K^>$. The smaller Kondo scale,
$T_K^<$,  gradually increases as we approach the transition point, and therefore the conductance
decreases $T_K^<$ becomes larger than $B$, thus giving a dip on the triplet side of the transition.

The third case is very different. Here $B=0.1$ is larger than both Kondo scales, 
and therefore the bump on the triplet side is suppressed. Even the maximum associated with the 
triplet-singlet transition is shifted to larger values of $\Delta$. We suspect 
that in this case 
the bump has an entirely different origin and is due to the Kondo effect 
associated with the degeneracy of the $S_z = -1$ triplet state and the singlet 
state\cite{Zeeman}.

\section{Conclusions}
We have studied transport through a lateral quantum dot in the vicinity of 
a singlet--triplet transition in the ground state. 
Our calculation is based on the numerical renormalization group, which enables us 
to calculate the full many--body spectrum of the dot. 
To determine the conductance we used a new method, first outlined in 
[\onlinecite{Borda}],
where we extract the conduction electron phase shifts with high accuracy 
directly from the NRG spectrum and use them in combination 
with the Landauer formula to determine the conductance at zero temperature. 
The only requirement for this technique to work is that the system considered must have 
a Fermi liquid ground state with the phase shifts being well-defined quantities.

In our calculations have modeled the quantum dot by a two--level Anderson impurity Hamiltonian 
with asymmetric couplings to the left and right single--mode lead. 
Our results show that at $T=0$ the sharp quantum phase transition 
that occurs for symmetric couplings 
becomes a \emph{crossover} at finite asymmetry. As a function of level splitting, 
the conductance shows a characteristic maximum (``bump'') at the crossover point and decreases 
on both the singlet and the triplet side. 

In order to model the effect of finite temperature or bias voltage, we have 
applied an in--plane magnetic field $B$ leading to a Zeeman--split triplet. 
We find that on both the singlet and the triplet side, the conductance has 
a \emph{non--monotonic} behavior as a function of the Zeeman field, thus 
leading to a characteristic ``dip'' at low $B$. Because of the qualitative 
equivalence between $B$ and finite bias voltage, we expect this structure 
to appear in the differential conductance $dI/dV$ as well. 
The associated low--energy scale (dip width) increases linearly on the singlet side, 
where it corresponds to the renormalized singlet--triplet splitting. 
On the triplet side, it 
%vanishes
decreases  rapidly with increasing distance from the 
degeneracy point, which is in agreement with two--stage Kondo screening of the triplet. 

These findings are consistent with the experimental results of 
van der Wiel \emph{et al.}~[\onlinecite{vanderWiel02}], 
where a dip in $dI/dV$ was indeed found on both sides of the singlet--triplet transition. 
On the other hand, in the measurements of Kogan \emph{et al.}~[\onlinecite{Kogan02}], 
the dip structure was only observed on the singlet side, indicating 
a weak asymmetry below experimental resolution. 

We expect our results to be relevant also for transport through other multi--level 
Kondo systems like carbon nanotubes or molecules. 

Acknowledgements: 
We are grateful to M.~\mbox{Pustilnik}, L.~Glazman, M.~Kastner, G.~Granger 
and A.~Kogan for illuminating discussions.  
This research has been supported  by NSF Grants DMR-99-81283, %Halperin99 
DMR-02-33773, %Halperin02 
Hungarian Grants OTKA F030041,   %Gergely 
T038162, %Patrik
and  the EU 'Spintronics' Research Training Network. 
W.H. acknowledges financial support from the German Science Foundation (DFG).
G.Z. is a Bolyai fellow.


\begin{references}
\bibitem{large_N}
Strictly speaking, Eq.~(\ref{H_dot}) is fully justified only for very 
large dots with chaotic single-particle states\cite{ABG}. 
For ballistic motion of electrons within the dot the magnitude of the terms not 
included in Eq.~(\ref{H_dot}) is of the order of $N^{-1/4}\delta E$. Inclusion of 
these terms is not expected to change the results qualitatively\cite{PG2}.
\end{references}
\end{document}